\documentclass[12pt]{article}
\usepackage{graphics,epsfig}
\usepackage[english]{babel}
\newcommand{\cinst}[2]{$^{\mathrm{#1}}$~#2\par}
\newcommand{\crefi}[1]{$^{\mathrm{#1}}$}

\begin{document}


\thispagestyle{empty}
\begingroup




\begin{center}



\vglue 1.0cm {\Large{\bf A study of the double hadron
neutrinoproduction on nuclei}}

\end{center}

\vspace{1.cm}

\begin{center}
{\large SKAT Collaboration}

\vspace{1.cm}

 N.M.~Agababyan\crefi{1}, V.V.~Ammosov\crefi{2},
 M.~Atayan\crefi{3},\\
 L.~Grigoryan\crefi{3}, N.~Grigoryan\crefi{3}, H.~Gulkanyan\crefi{3}, \\
 A.A.~Ivanilov\crefi{2}, Zh.~Karamyan\crefi{3},
V.A.~Korotkov\crefi{2}

\small

\vspace{1.cm} \cinst{1}{Joint Institute for Nuclear Research,
Dubna, Russia} \cinst{2}{Institute for High Energy Physics,
Protvino, Russia} \cinst{3}{Yerevan Physics Institute, Armenia}
\end{center}

\vspace{100mm}

{\centerline{\bf YEREVAN  2006}}


\newpage
\vspace{1.cm}
\begin{abstract}
The nuclear medium influence on the dihadron neutrinoproduction is
investigated for the first time, using the data obtained with SKAT
bubble chamber. An indication is obtained that the nuclear
attenuation of the dihadron is more expressed for kinematically
closest hadron pairs. The experimental data on the dihadron
attenuation and on the ratio of the dihadron to single-hadron
yields are compared with predictions of the two-scale string
fragmentation model.
\end{abstract}

\newpage
\setcounter{page}{1}
\section{Introduction}


The space-time structure of the quark string fragmentation in
leptoproduction reactions imposes certain correlations between
two-hadron formation processes. A unique information on the latter
can be inferred from experiments on nuclear targets
(\cite{ref1,ref2,ref3,ref4,ref5} and references therein). The
first measurements of the dihadron leptoproduction, performed
recently \cite{ref6} using a 27.6 GeV positron beam with
deuterium, nitrogen, krypton and xenon targets, showed that the
available theoretical models do not satisfactorily describe the
experimental data, especially on heavier targets.\\ The aim of the
present work is to obtain the first experimental data on the
dihadron neutrinoproduction at comparatively lower energies, at
which the nuclear attenuation effects are expected to be more
prominent. To this end, the data from the SKAT bubble chamber
\cite{ref7} were used. In Section 2, the experimental procedure is
described. Section 3 presents the data on the dihadron attenuation
versus its kinematical variables. In Section 4, the data on the
dihadron attenuation are compared with those for the single hadron
and with theoretical predictions. The results are summarized in
Section 5.

\section{Experimental procedure}

\noindent The experiment was performed with SKAT bubble chamber,
exposed to a wideband neutrino beam obtained with a 70 GeV primary
protons from the Serpukhov accelerator. The chamber was filled
with a propane-freon mixture containing 87 vol\% propane
($C_3H_8$) and 13 vol\% freon ($CF_3Br$) with the percentage of
nuclei H:C:F:Br = 67.9:26.8:4.0:1.3 \%. A 20 kG uniform magnetic
field was provided within the operating chamber volume.
\\ Charged current interactions containing a negative muon with momentum
$p_{\mu} >$0.5 GeV/c were selected. Other negatively charged
particles were considered to be $\pi^-$ mesons. Protons with
momentum below 0.6 GeV$/c$ and a fraction of protons  with
momentum 0.6-0.85 GeV$/c$ were identified by their stopping in the
chamber. Non-identified positively charged particles were
considered to be ${\pi}^+$ mesons. Events in which errors in
measuring the momenta of all charged secondaries and photons were
less than 27\% and 100\%, respectively, were selected. Each event
is given a weight which corrects for the fraction of events
excluded due to improperly reconstruction. More details concerning
the experimental procedure, in particular, the reconstruction of
the neutrino energy $E_{\nu}$ can be found in our previous
publications \cite{ref8,ref9}. \\ The events with $3 < E_{\nu} <$
30 GeV were accepted, provided that the reconstructed mass $W$ of
the hadronic system exceeds 2 GeV and $y = \nu/E_\nu < 0.95$,
$\nu$ being the energy transferred to the hadronic system. No
restriction was imposed on the transfer momentum squared $Q^2$.
The number of accepted events was 3440 (4819 weighted events). The
mean values of the kinematical variables were $<E_{\nu}>$ = 10.0
GeV, $<W>$ = 2.9 GeV, $<W^2>$ = 9.1 GeV$^2$, $<Q^2>$ = 2.7
(GeV/c)$^2$ and $<\nu>$ = 5.8 GeV. \\ Further, the whole event
sample was subdivided, using several topological and kinematical
criteria \cite{ref9,ref10}, into three subsamples: the 'cascade'
subsample $B_S$ with a sign of intranuclear secondary interaction,
the 'quasiproton' ($B_p$) and 'quasineutron' ($B_n$) subsamples.
About 40\% of subsample $B_p$ is contributed by interactions with
free hydrogen. Weighting the 'quasiproton' events with a factor of
0.6, one can compose a 'pure' nuclear subsample $B_A = B_S + B_n +
0.6 B_p$ and a 'quasinucleon' subsample $B_N = B_n + 0.6 B_p$. It
has been verified \cite{ref9,ref10,ref11}, that the multiplicity
and spectral characteristics of secondary particles in the $B_p
(B_N)$ subsample are in satisfactory agreement with those measured
with a pure proton (deuteron) target. The effective atomic weight
corresponding to the subsample $B_A$ is estimated \cite{ref12} to
be approximately equal to $A_{eff} = 21 \pm 2$, when taking into
account the probability of secondary intranuclear interactions in
the composite target.

\section{The nuclear attenuation versus dihadron kinematical
variables}

Below we will consider hadrons produced in the forward hemisphere
in the hadronic c.m.s. (i.e. in the region of $x_F > 0$, $x_F$
being the Feynman variable), because in this region the nuclear
attenuation effects for hadrons dominate over intranuclear
cascading effects which lead to an enhancement of their yield,
observable in the region of $x_F < 0$ (see e.g.
\cite{ref9,ref11}).
\\ In this section, we present the data which demonstrate the
nuclear medium influence on the dihadron yield in dependence on
kinematical variables describing the double-hadron system. For a
given variable $v$, this influence is characterized by the ratio
$R_2(v)$ of the dihadron differential yields $<n_2(v)>_A$ and
$<n_2(v)>_N$ in the nuclear and 'quasinucleon' subsamples,
$R_2(v)=<n_2(v)>_A/<n_2(v)>_N$. \\ Figs. 1-3 demonstrate the
dependence of $R_2$ on the relative kinematical variables of two
hadrons: $\Delta{x_F}=|x_{F1}-x_{F2}|$; $\Delta{z}=|z_1-z_2|$,
$z_1$ and $z_2$ being the fractions of the quark energy $\nu$
carried by hadrons; $\Delta{y}^*=|y_1-y_2|$, $y_1$ and $y_2$ being
the rapidities in the hadronic c.m.s.;
$\eta_{rel}=-\ln[\tan(\vartheta_{rel}/2)]$, $\vartheta_{rel}$
being the angle between two hadrons;
$Q_{inv}=\sqrt{-(p_1-p_2)^2}$, $p_1$ and $p_2$ being the
four-momenta of hadrons;
$q_t=|\overrightarrow{p_{t1}}-\overrightarrow{p_{t2}}|$,
$\overrightarrow{p_{t1}}$ and $\overrightarrow{p_{t2}}$ being the
transverse momenta of hadrons relative to the current quark
direction. The data are presented separately for the like-sign
($\pi^+\pi^+$), unlike-sign ($\pi^+\pi^-$) and charged ($\pi\pi$)
hadron pairs. \\ The most interesting feature of the data of Figs.
1-3 is that the attenuation of the unlike-sign pair produced in
the quark fragmentation region (at $x_F > 0.15$) tends to
strengthen with decreasing relative kinematical variables (the
middle panels of figures); the said also concerns the relative
angle $\vartheta_{rel}$ (which decreases with increasing
$\eta_{rel}$). This tend, as it can be seen from figures, is not
caused by the $\rho^0$ meson production. The said feature is less
expressed (or even absent) for the like-sign pairs (the top panels
of Figs. 1-3). Note also, that no definite dependence of $R_2$ on
the relative kinematical variables is observed at a less strict
cut on $x_F (x_F > 0)$. \\ Figs. 4 and 5 show the ratio $R_2$
versus the collective kinematical variables of the dihadron:
$x_F^{pair}$, $z^{pair}$, the rapidity $y^{pair}$ and the
transverse momentum $p_t^{pair}$. For the case of $x_F > 0.15$, no
significant dependence on collective variables is observed, except
a possible rise of $R_2$ with increasing $z^{pair}$ (the right
panel of Fig. 4). For the case of the cut $x_F > 0$, the value
$R_2$ is consistent with 1 at the smallest value of $x_F^{pair}$,
$z^{pair}$, and $y^{pair}$, while at larger values of the latters
a significant attenuation $(R_2 < 1)$ is observed. An indication
is obtained, that the attenuation of the like-sign pair at $x_F >
0$ weakens with increasing $p_t^{pair}$ (Fig. 5, the right top
panel).

\section{The dihadron versus single-hadron attenuation}

The correlation between the two-hadron attenuation as a function
of  certain kinematical variable $v$ can be characterized by the
double ratio

\begin{equation}
r_{21}(v_1,v_2) \, = \, \frac{R_2(v_1,v_2)}{R_1(v_1)}
\end{equation}

\noindent where the ratio $R_1(v)=<n_1(v)>_A/<n_1(v)>_N$ is the
single-hadron attenuation factor. \\ It should be stressed here,
that even at the absence of dinamical correlations between
attenuations of two hadrons, the paths they pass in the nuclear
medium (and hence their absorption probabilities) are always
'geometrically' correlated, resulting in a certain dependence of
$r_{21}(v_1,v_2)$ on $v_2$ (at fixed $v_1$). If the nuclear
attenuation is dominated by the absorption of the (pre)hadronic
states, then the double ratio $r_{21}(v_1,v_2)$ is always
obviously smaller than 1. Below we will consider the dependence of
$r_{21}(z_1,z_2)$ on variables $z_1$ and $z_2$ of two hadrons.
\\ Before presenting the data on the double ratio, it will be
useful to consider separately the single-hadron and dihadron
attenuation factors $R_1(z_1)$ and $R_2(z_1,z_2)$. The latters are
plotted in Figs. 6 and 7 for charged hadrons with $x_F > 0$ and
$x_F > 0.1$, respectively. The values of $R_1(z_1)$ are quoted
numerically and indicated, for visual comparison with
$R_2(z_1,z_2)$, by dashed lines. As it is seen, in general
$R_2(z_1,z_2) < R_1(z_1)$ due to the additional attenuation of the
yield of the accompanying hadron with $z=z_2$.
\\  As it was already mentioned in the previous section, the
attenuation is more expressed for kinematically close pairs, in
particular, at $z_1$, $z_2 < 0.12$ and $0.24 < z_1$, $z_2 <0.37$
(Fig. 6) and $0.12 < z_1$, $z_2 < 0.24$ and $0.24 < z_1$, $z_2 <
0.34$ (Fig. 7). Similar observations can be done from Fig. 8 and
9, where the double ratio $r_{21}(z_1,z_2)$ is plotted as a
function of $z_1$ and $z_2$ for charged hadrons with $x_F > 0$ and
$x_F > 0.1$, respectively.
\\ The double ratio $r_{21}(z_1,z_2)$ was recently measured in
electronuclear interactions \cite{ref4}, at $z_1=z_{tr} >$ 0.5 for
the leading trigger particle and $z_2=z_{sub}$ for the subleading
particle, being the fastest one among particles accompanying the
leading particle. Fig. 10 presents the data of \cite{ref4} on the
nitrogen target $(A=14)$ for charged pion combinations (the top
panel) and, in the bottom panel, for all pion combinations
(including $\pi^0$ mesons), except $\pi^+\pi^-$, in order to
exclude the contribution from the $\rho^0$ decay. In the same
figure we also plot our data for charged pion combinations (the
top panel), for combinations including also $\pi^0$ mesons (the
middle panel), and the same but without $\pi \pi$ combinations
from the $\rho$ mass region. As it is seen, our data are mainly
compatible with electroproduction data, exhibiting, however, an
enhancement at the low- $z_2$ region, probably due to the more
pronounced intranuclear cascading effects in a  slightly heavier
composite target ($A_{eff} \approx 21$) used in our experiment. \\
We undertook an attempt to describe our data on $R_1(z)$,
$R_2(z_1,z_2)$ and the double ratio $r_{21}(z_1,z_2)$ in the
framework of the Two-Scale Model (TSM) \cite{ref13} (see also
\cite{ref5,ref14}) which incorporates the space-time pattern of
the quark string fragmentation followed from the Lund model
\cite{ref1,ref15}. The TSM contains four free parameters: the
quark string tension $\kappa$ and the following string-nucleon
interaction cross sections, namely, $\sigma_q$ for the initial
string stretched between the struck quark and the target nucleon
remnant, $\sigma_s$ for the intermediate string stretched between
the struck quark and a created antiquark (which becomes a valence
one for the hadron being looked at) and $\sigma_h$ for the formed
colorless system with quantum numbers and valence content of the
final hadron.
\\ First of all we checked the compatibility of our data on the
single hadron attenuation factor with the TSM. A satisfactory
description of the data on $R_1(z)$ for hadrons with $x_F > 0.1$
(quoted in Fig. 7), as well as for hadrons with $x_F >0.1$ and $z
> 0.5$ (quoted in Fig. 6, the left bottom panel) can be reached at
the following set of the model parameters: $\kappa$ = 0.8 GeV/fm,
$\sigma_q = 0, \sigma_s$ =10mb and $\sigma_h$ = 20mb (note, that
this set somewhat differs from those estimated in \cite{ref13} and
\cite{ref14}). The TSM predictions for $R_2(z_1,z_2)$ and
$r_{21}(z_1,z_2)$, obtained with these parameters, are plotted in
Figs. 7, 9 and 10. As it is seen, the model predictions are, in
general, compatible with experimental data. The model, however,
does not describe the reduced yield of the pair of low-$z$ hadrons
with $0.12 < z_1, z_2 < 0.24$ (Figs. 7 and 9, the left top panel).
A similar inconsistency is also seen for $R_2(\Delta z)$ at low
values of variable $\Delta z = |z_1 - z_2|$ for charged hadrons
with $x_F > 0.15$ (Fig. 1, the right bottom panel). On the other
hand, the model underestimates the value of $R_2$ ($z^{pair}$) at
largest values on the collective variable $z^{pair} = z_1 + z_2$
for charged hadrons with $x_F > 0.1$ (Fig. 4, the right bottom
panel). Note also, that the data on the double ratio $r_{21} (z_1
> 0.5, z_2)$ at $z_2 < 0.24$ (Fig. 10, the bottom panel)
significantly exceed the model predictions, probably due to the
fact that the latter do not incorporate the intranuclear cascading
processes which could enhance the yield of low- $z$ hadrons, as
well as due to a possible contamination of the data by
non-identified protons.

\section{Summary}

For the first time for neutrinonuclear interactions, the nuclear
medium influence on the dihadron production is investigated, at an
effective atomic weight $A_{eff} \approx 21$ of a composite
target. \\ The dihadron attenuation factor, $R_2$, is measured
versus various kinematical veriables of the dihadron. An
indication is obtained, that the nuclear attenuation strengthens
with decreasing relative variables of two hadrons, reaching about
$R_2 \approx 0.7$ for kinematically closest pairs. \\ It is shown,
that the experimental data on the attenuation factor
$R_2(z_1,z_2)$ and the double ratio $r_{21}(z_1,z_2) =
R_2(z_1,z_2)/R_1(z_1)$ are compatible with predictions of the
two-scale string fragmentation model, except for the case of pairs
of low-$z$ hadrons with $0.12 < z_1$, $z_2 < 0.24$ and pairs of a
low-$z$ ($z < 0.24$) subleading and a high-$z$ ($z > 0.5$) leading
hadrons. \\

{\bf Acknowledgement.} The authors from YerPhI acknowledge the
supporting grants of Calouste Gulbenkian Foundation and Swiss
Fonds "Kidagan". The activity of two of the authors (L.G. and
H.G.) is supported by Cooperation Agreement between DESY and
YerPhI signed on December 6, 2002.


\newpage
\begin{figure}[ht]
 \resizebox{1.0\textwidth}{!}{\includegraphics*[bb =20 80 600
700]{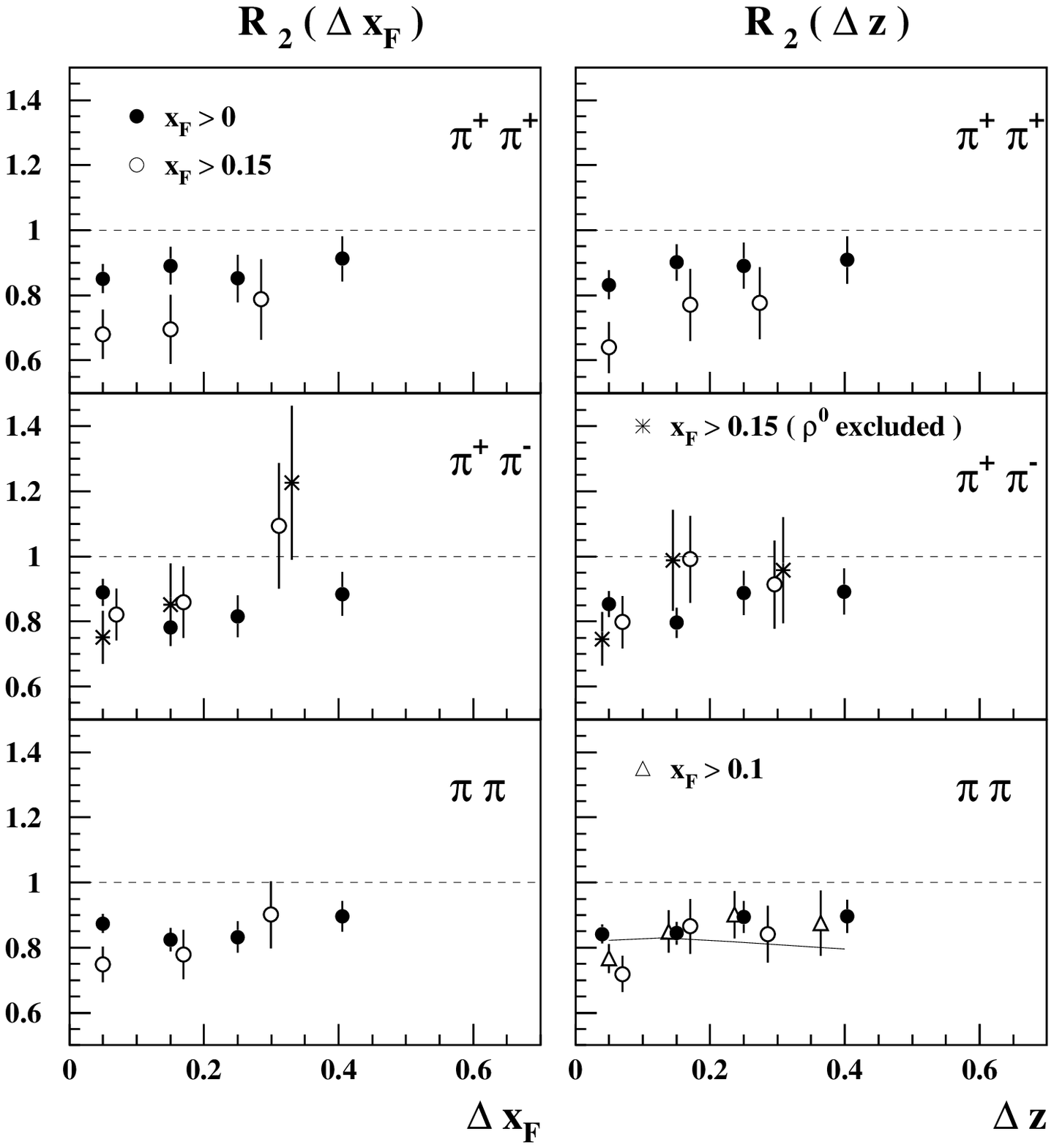}} \caption{The dependence of the dihadron
attenuation factor $R_2$ on $\Delta {x_F}$ (left panel) and
$\Delta z$ (right panel). The curve is the model prediction (see
text).}
\end{figure}

\newpage
\begin{figure}[h]

\resizebox{0.8 \textwidth}{!}{\includegraphics*[bb=65 60 500 650]
{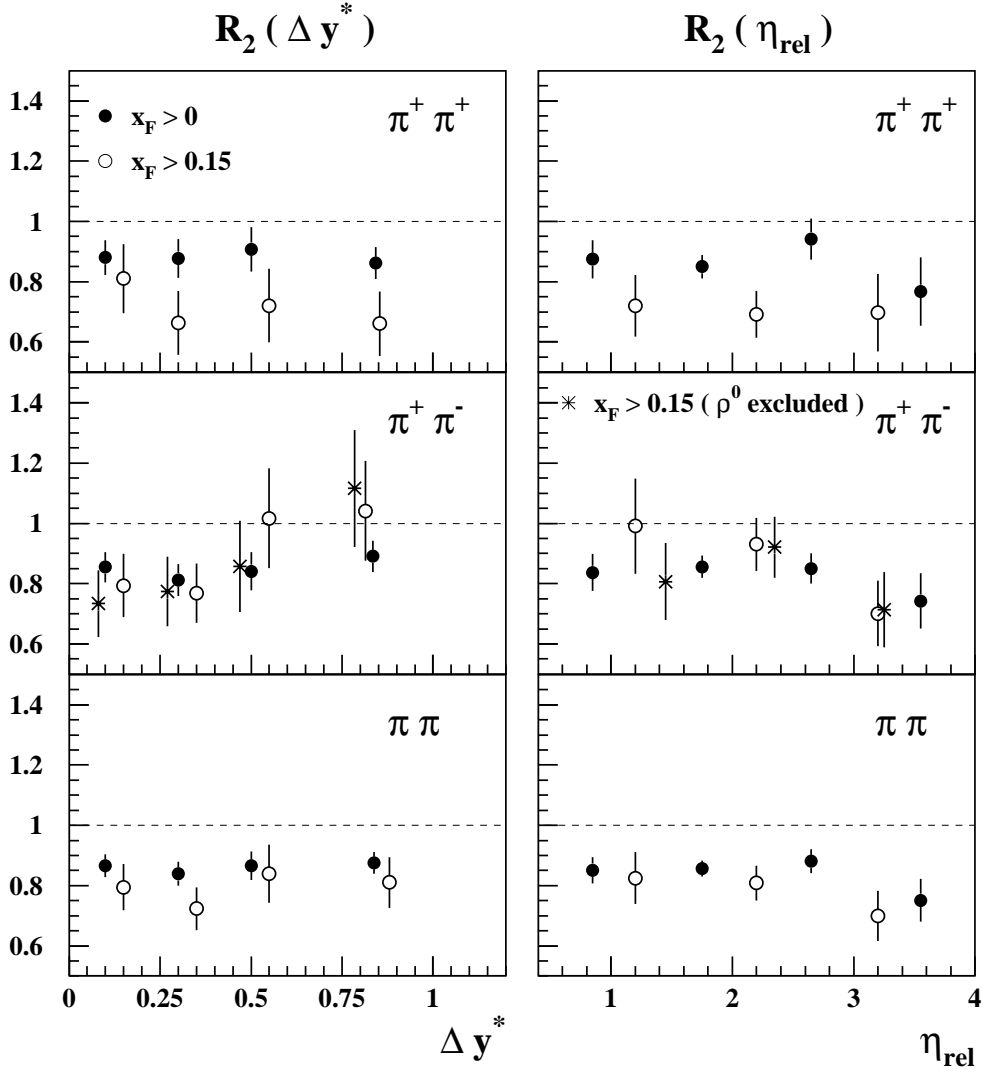}} \caption{ The dependence of the dihadron attenuation
factor $R_2$ on $\Delta {y^*}$ (left panel) and $\eta_{rel}$
(right panel).}
\end{figure}

\newpage
\begin{figure}[h]
\resizebox{0.8 \textwidth}{!}{\includegraphics*[bb=65 60 500
650]{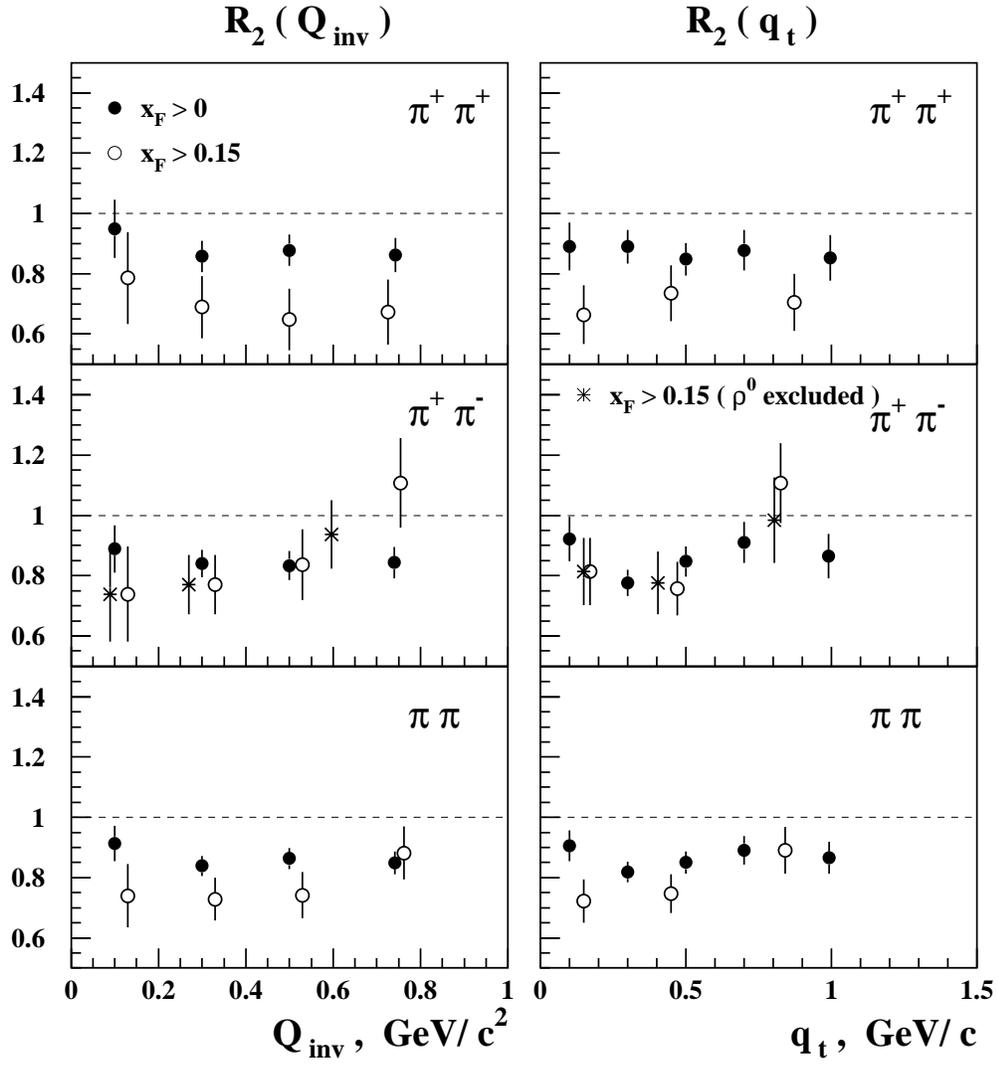}} \caption{The dependence of the dihadron
attenuation factor $R_2$ on $Q_{inv}$ (left panel) and $q_t$
(right panel).}
\end{figure}

\newpage
\begin{figure}[ht]
\resizebox{0.8 \textwidth}{!}{\includegraphics*[bb=65 60 500
650]{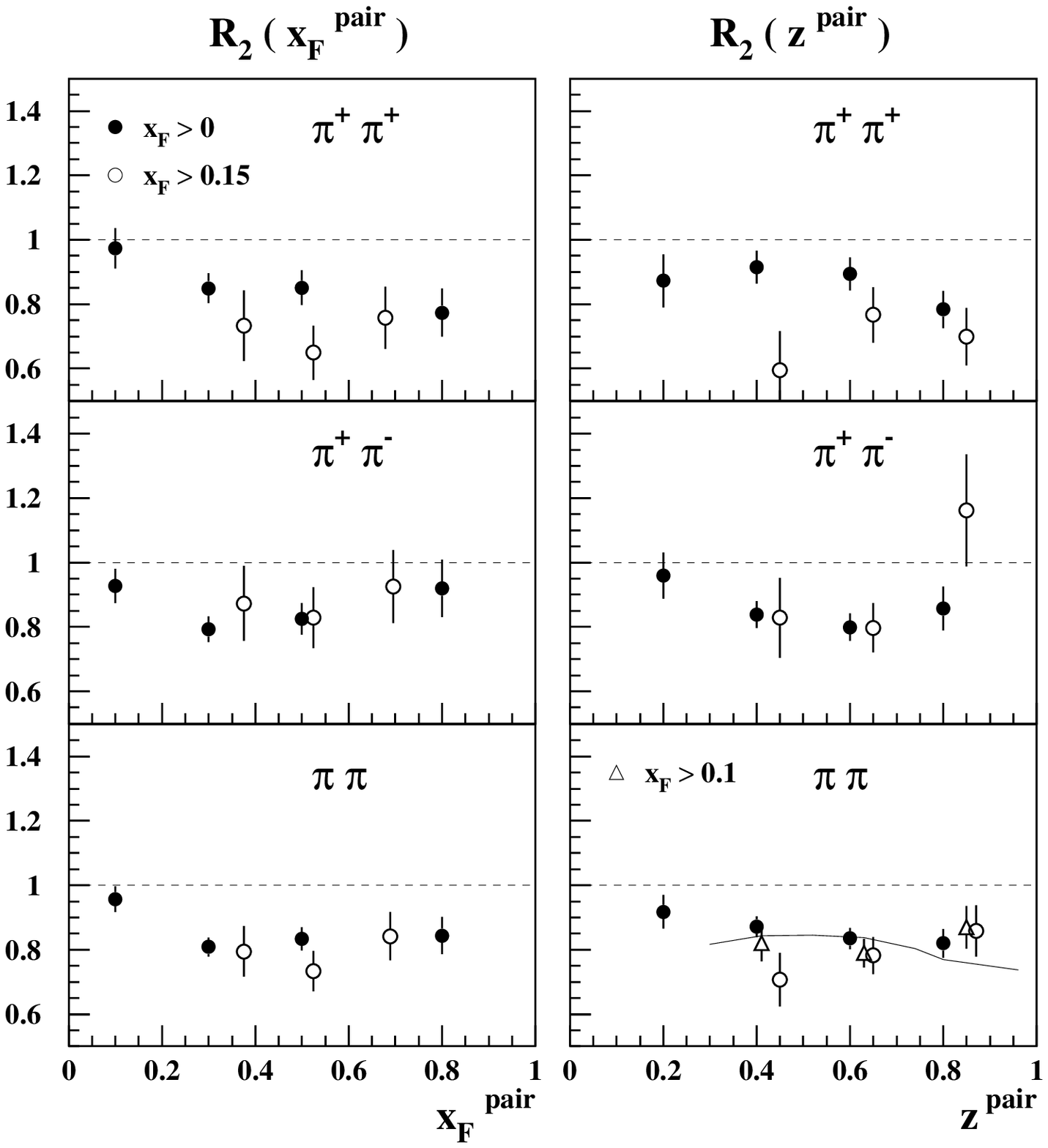}} \caption{The dependence of the dihadron
attenuation factor $R_2$ on $x_F^{pair}$ (left panel) and
$z^{pair}$ (right panel). The curve is the model prediction (see
text).}
\end{figure}

\newpage
\begin{figure}[ht]
\resizebox{0.8 \textwidth}{!}{\includegraphics*[bb=65 60 500
650]{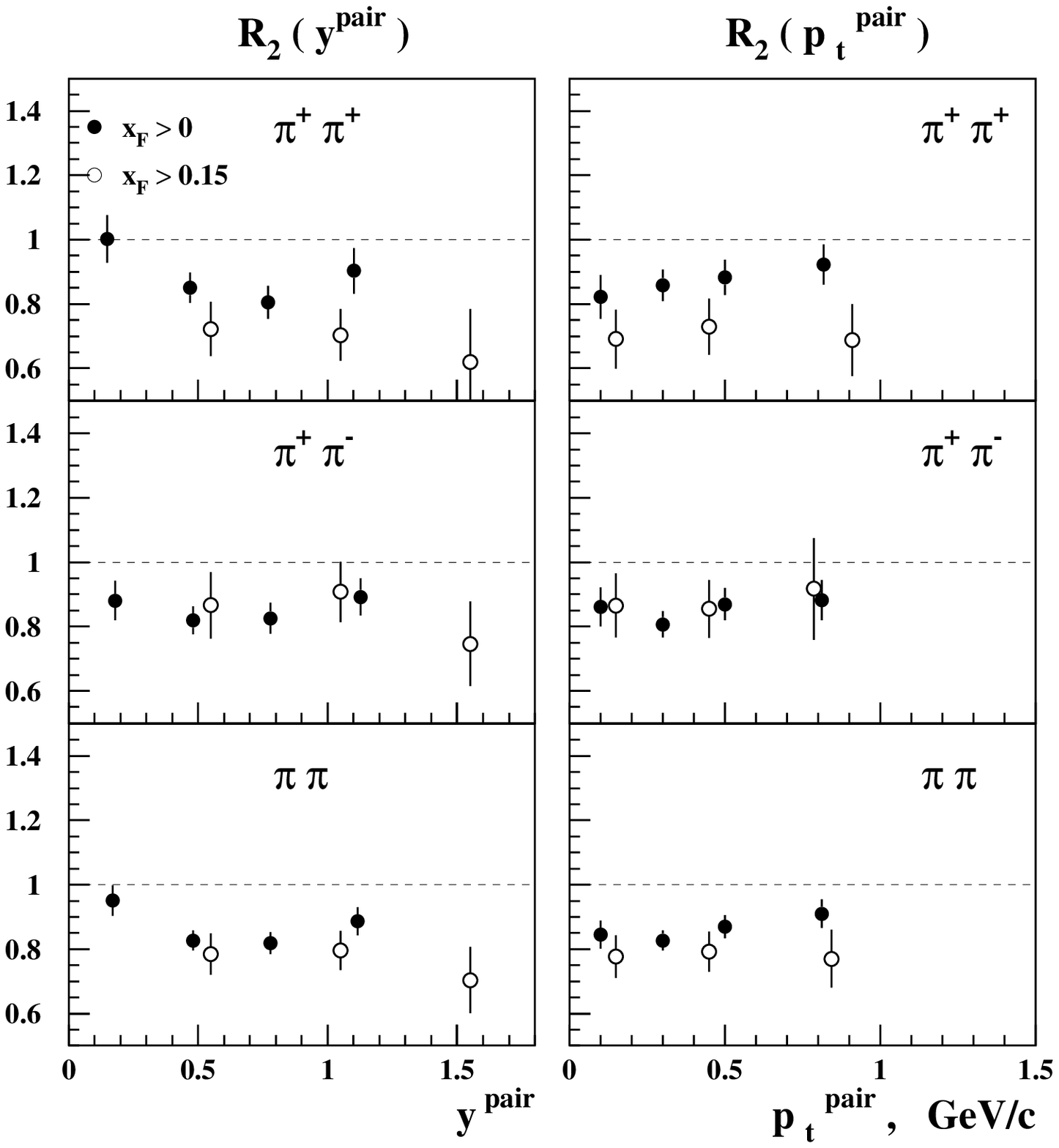}} \caption{The dependence of the dihadron
attenuation factor $R_2$ on $y^{pair}$ (left panel) and
$p^{pair}_t$ (right panel).}
\end{figure}

\newpage
\begin{figure}[ht]
\resizebox{0.8 \textwidth}{!}{\includegraphics*[bb=65 80 500
650]{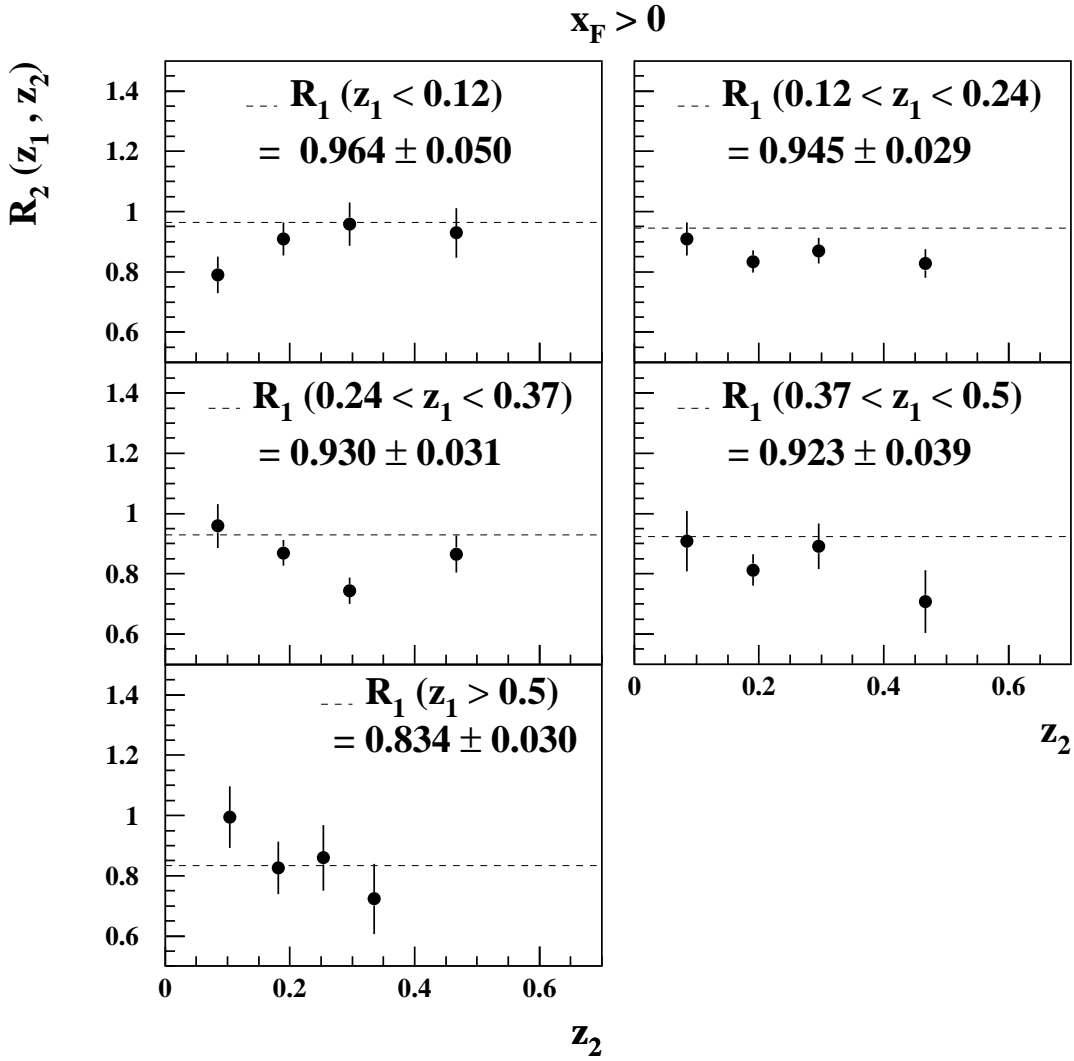}} \caption{The dependence of the dihadron
attenuation factor $R_2(z_1,z_2)$ as a function of $z_1$ (for
'trigger' particle) and $z_2$ (for accompanying particle). The
numerical values and the dashed lines in figures concern the
single-particle attenuation factor $R_1(z_1)$. The cut $x_F > 0$
was applied.}
\end{figure}

\newpage
\begin{figure}[ht]
\resizebox{0.8 \textwidth}{!}{\includegraphics*[bb=65 60 500
650]{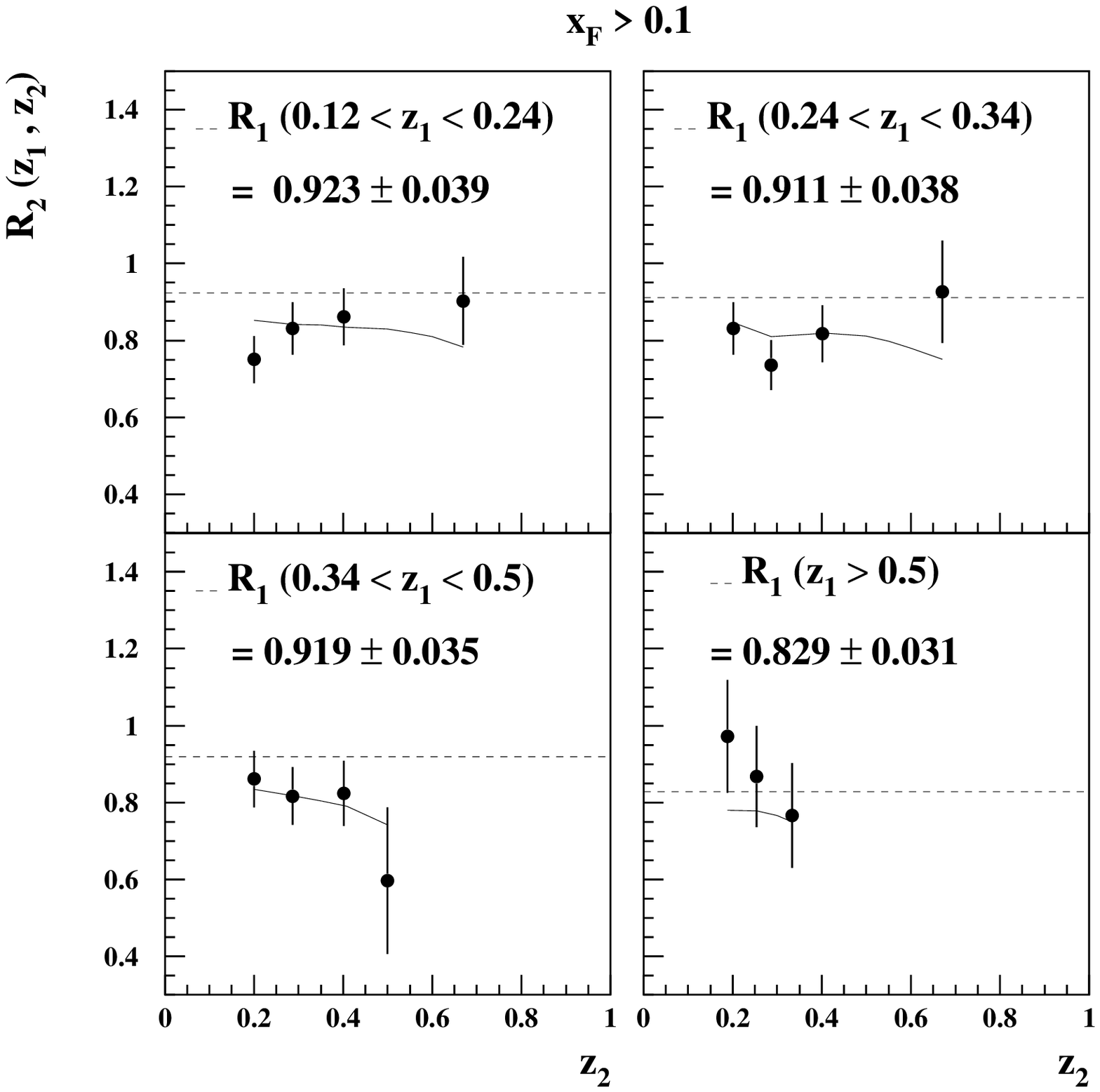}} \caption{The same as Figure 6, but the cut $x_F >
0.1$ is applied. The curves are predictions of the two-scale
string fragmentation model.}
\end{figure}

\newpage
\begin{figure}[ht]
\resizebox{0.8 \textwidth}{!}{\includegraphics*[bb=10 80 500
750]{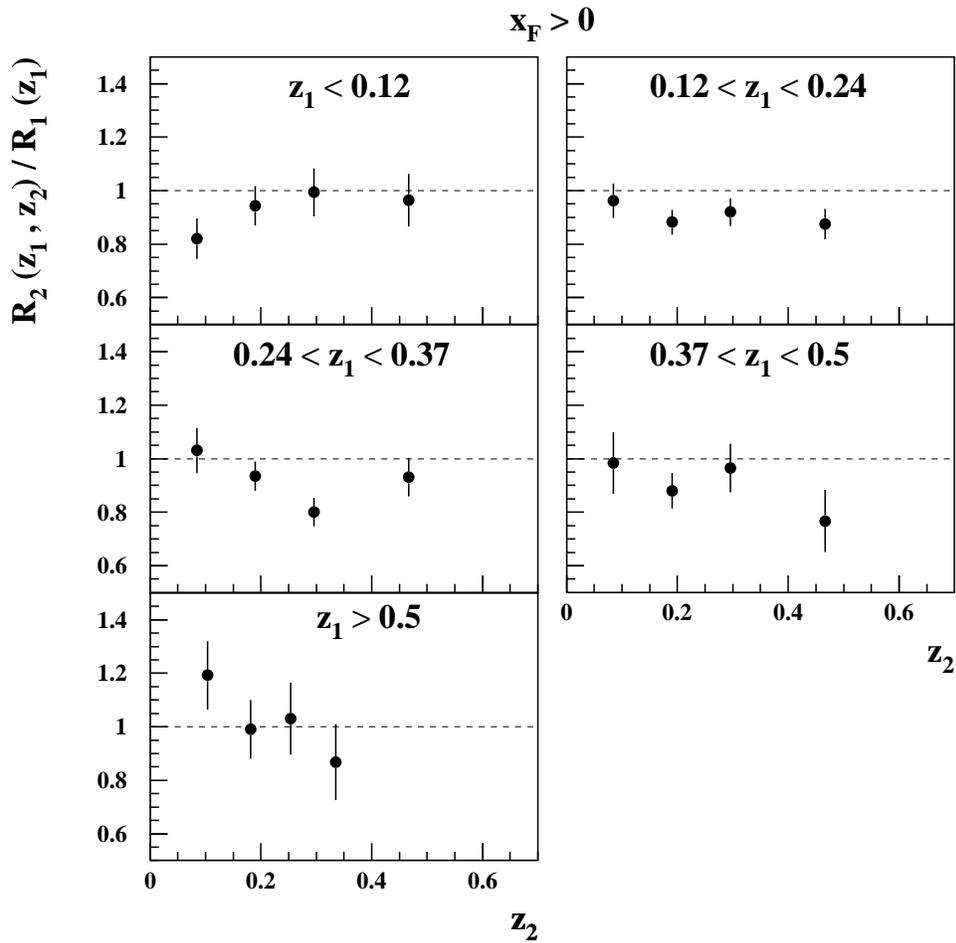}} \caption{The double ratio
$r_{21}(z_1,z_2)=R_2(z_1,z_2)/R_1(z_1)$ as a function of $z_1$ and
$z_2$. The cut $x_F > 0$ was applied.}
\end{figure}

\newpage
\begin{figure}[ht]
\resizebox{0.8 \textwidth}{!}{\includegraphics*[bb=10 80 500
750]{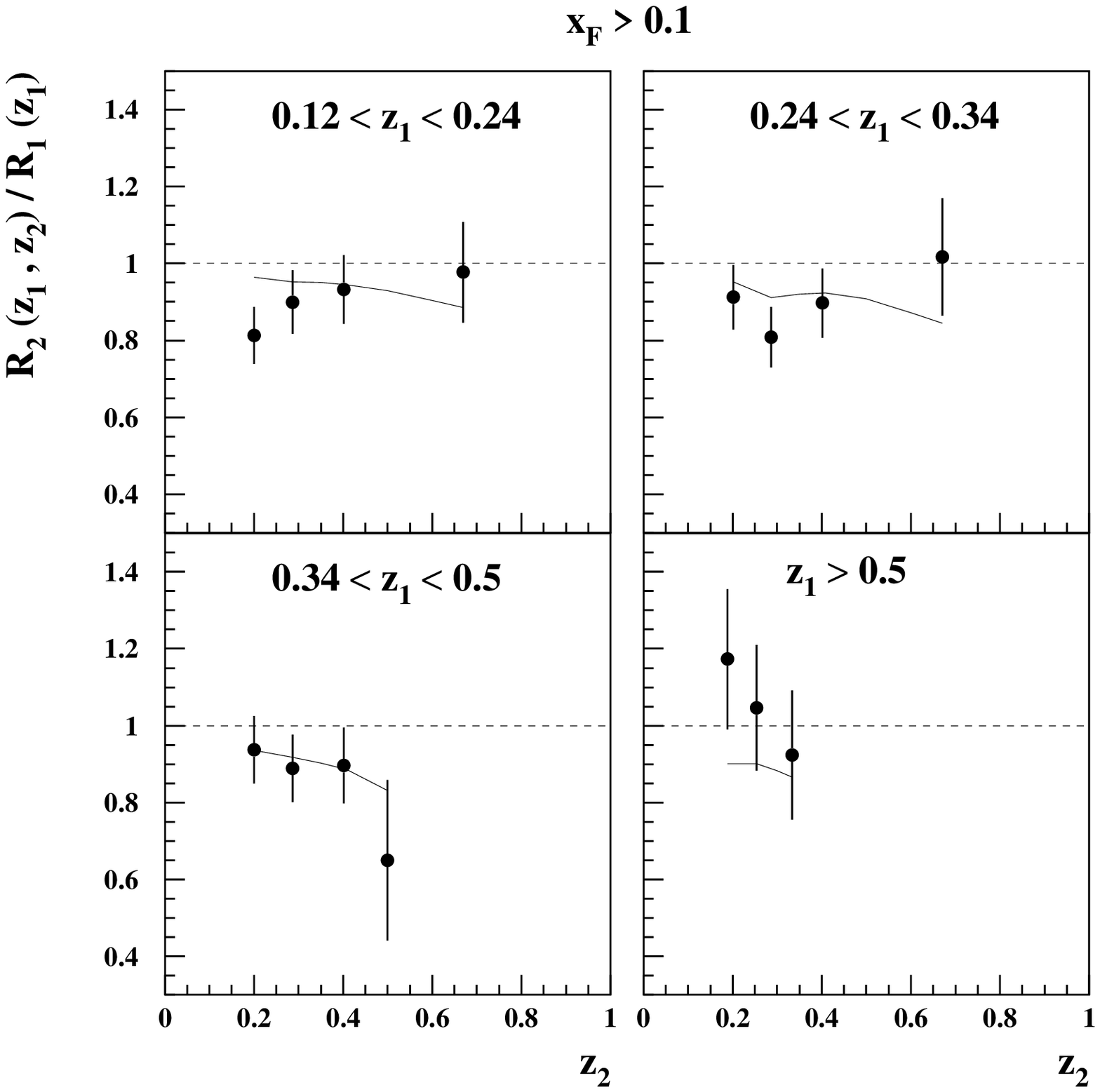}} \caption{The same as Figure 8, but the cut $x_F >
0.1$ is applied. The curves are predictions of the two-scale
string fragmentation model.}
\end{figure}

\newpage
\begin{figure}[ht]
\resizebox{0.8 \textwidth}{!}{\includegraphics*[bb=10 100 500
750]{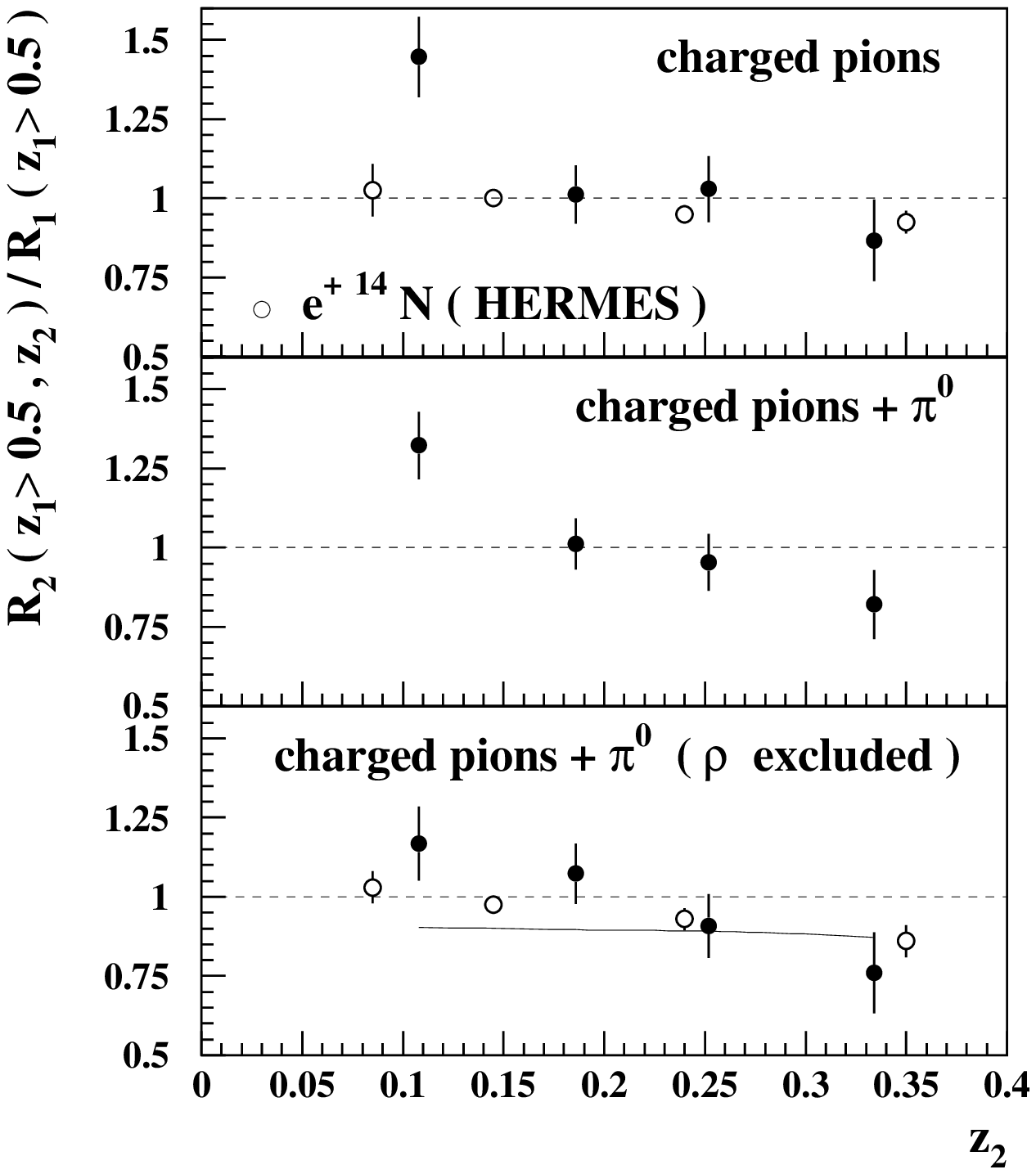}} \caption{The double ratio $r_{21}(z_1 > 0.5,z_2)$
as a function of $z_2$ at $z_1 > 0.5$. The cut $x_F > 0$ was
applied. The curve is prediction of the two-scale string
fragmentation model.}
\end{figure}


\begin{thebibliography}{00}
\parskip=0.pt \parsep=0.pt \itemsep=0.pt
\bibitem{ref1}
J.Czy\u{z}ewski, Phys. Rev. {\bf{C 43}}, 2426 (1991)
\bibitem{ref2}
B.Z.Kopeliovich et al., Nucl. Phys. {\bf{A 740}}, 211 (2004)
\bibitem{ref3}
A.Majumder, nucl-th/0503019
\bibitem{ref4}
T.Falter, K.Gallmeister, U.Mozel, nucl-th/0509041
\bibitem{ref5}
N.Akopov, L.Grigoryan, Z.Akopov, hep-ph/0605128
\bibitem{ref6}
A.Airapetian et al. (HERMES Coll.), Phys. Rev. Lett. {\bf96},
162301 (2006)
\bibitem{ref7}
V.V.Ammosov et al. Fiz. Elem. Chastits At. Yadra {\bf23}, 648,
1992 [Sov. J. Part. Nucl. {\bf23}, 283, (1992)]
\bibitem{ref8}
N.M.Agababyan et al. (SKAT Coll.), YerPhI Preprint N 1535
(Yerevan, 1999)
\bibitem{ref9}
N.M.Agababyan et al. (SKAT Coll.), Yad. Fiz. {\bf66}, 1350 (2003)
[Phys. of At. Nucl. {\bf66}, 1310 (2003)]
\bibitem{ref10}
N.M.Agababyan et al. (SKAT Coll.),YerPhI Preprint N 1578 (Yerevan,
2002)
\bibitem{ref11}
N.M.Agababyan et al. (SKAT Coll.), Yad. Fiz. {\bf68}, 1241 (2005)
[Phys. of At. Nucl. {\bf 68}, 1160 (2005)]
\bibitem{ref12}
N.M.Agababyan et al. (SKAT Coll.),hep-ex/0504024; YerPhI Preprint
N 1597 (Yerevan, 2005);
\bibitem{ref13}
J.Ashman et al. Z. Phys. {\bf{C 52}}, 1 (1991)
\bibitem{ref14}
N.Akopov, L.Grigoryan, Z.Akopov, Eur. Phys. J. {\bf{C 44}}, 219
(2005)
\bibitem{ref15}
A.Bialas, M.Gyulassy, Nucl. Phys. {\bf {B 291}}, 793 (1987)
\bibitem{ref16}
A.Airapetian et al. (HERMES Coll.), Eur. Phys. J. {\bf {C 20}},
479 (2001)

\end{thebibliography}
\end{document}